\documentclass[useAMS,usenatbib]{mn2e}
\usepackage{ amssymb }
\usepackage{amsmath}
\usepackage{graphicx}
\usepackage{color}

\title[Clustering tomography]{Clustering tomography: measuring cosmological distances through angular clustering in thin redshift shells}
\author[S. Salazar-Albornoz, et al.]{\parbox[t]{\textwidth}{\vspace{-1cm}
Salvador Salazar-Albornoz$^{1,2,3}$\thanks{E-mail: ssalazar@mpe.mpg.de}, 
Ariel G. S\'anchez$^2$, Nelson D. Padilla$^{3,4,5}$ \\and Carlton M. Baugh$^{6}$}\\
$^1$Universit\"ats-Sternwarte M\"unchen, Scheinerstrasse 1, 81679 Munich, Germany\\
$^2$Max-Planck-Institut f\"ur Extraterrestrische Physik, Giessenbachstrasse 1, 85748 Garching, Germany\\
$^3$Instituto de Astrof\'isica, Pontificia Universidad Cat\'olica de Chile, Santiago, Chile\\
$^4$Centro de Astro-Ingenier\'ia, Pontificia Universidad Cat\'olica de Chile, Santiago, Chile\\
$^5$Max Planck Institut f\"ur Astrophysik, Karl-Schwarzschild-Str. 1, D-85741, Garching, Germany\\
$^6$Institute for Computational Cosmology, Department of Physics, University of Durham, South Road, Durham DH1 3LE, UK}
\begin{document}

\date{Accepted 2014 July 14. Received 2014 July 14; in original form 2014 February 14}

\pagerange{\pageref{firstpage}--\pageref{lastpage}} \pubyear{2013}

\maketitle

\label{firstpage}

\begin{abstract}
We test the cosmological implications of studying galaxy clustering using a tomographic approach, by computing the galaxy two-point angular correlation function $\omega(\theta)$ in thin redshift shells using a spectroscopic-redshift galaxy survey. The advantages of this procedure are that it is not necessary to assume a fiducial cosmology in order to convert measured angular positions and redshifts into distances, and that it gives several (less accurate) measurements of the angular diameter distance $D_\rmn{A}(z)$ instead of only one (more precise) measurement of the effective average distance $D_\rmn{V}(z)$, which results in better constraints on the expansion history of the Universe. We test our model for $\omega(\theta)$ and its covariance matrix against a set of mock galaxy catalogues and show that this technique is able to extract unbiased cosmological constraints. Also, assuming the best-fit $\Lambda$CDM cosmology from the cosmic microwave background measurements from the Planck satellite, we forecast the result of applying this tomographic approach to the final Baryon Oscillation Spectroscopic Survey catalogue in combination with Planck for three flat cosmological models, and compare them with the expected results of the isotropic baryon acoustic oscillation (BAO) measurements post-reconstruction on the same galaxy catalogue combined with Planck. While BAOs are more accurate for constraining cosmological parameters for the standard $\Lambda$CDM model, the tomographic technique gives better results when we allow the dark energy equation of state $w_{DE}$ to deviate from $-1$, resulting in a performance similar to BAOs in the case of a constant value of $w_{DE}$, and a moderate improvement in the case of a time-dependent value of $w_{DE}$, increasing the value of the Figure-of-Merit in the $w_0-w_a$ plane up to $15\%$.
\end{abstract}


\section{Introduction}
The study of large scale structure (LSS) has been of great importance for the advancement of our understanding of the Universe, characterising the distribution of structures, such as galaxies and voids, at large scales. Supported by the increasing amount of data from current and future large galaxy surveys, such as the Baryon Oscillation Spectroscopic Survey \citep[BOSS; ][]{BOSS}, WiggleZ \citep{WggZ}, the Dark Energy Survey \citep[DES; ][]{DES}, the Hobby-Eberly Telescope Dark Energy Experiment \citep[HETDEX; ][]{HETDEX}, the Large Synoptic Survey Telescope \citep[LSST; ][]{LSST} and the EUCLID mission \citep{EUCLID}, in combination with new and more precise measurements of the cosmic microwave background (CMB), the study of the LSS has a promising future in terms of shedding light on the nature of the Universe.

One of the most important cosmological probes of LSS is the signal of the baryon acoustic oscillations (BAO) measured in two-point statistics, such as the correlation function or the power spectrum. These oscillations occur because small primordial perturbations induce sound waves in the relativistic plasma of the early Universe \citep{PeeblesYu70}, but later on at the recombination epoch ($z\approx1000$), the sound speed suddenly decreases to the point that these waves stop propagating. Since the Universe has an appreciable fraction of baryons, cosmological theories predict that the BAO signal will also be imprinted onto the two-point statistics of the non-relativistic matter as an excess of clustering in the correlation function, or an oscillation in power in the power spectrum, at a given (fixed) scale, making it an ideal standard ruler.

In 1999, motivated by the results obtained from COBE of the primary temperature anisotropy in the CMB \citep{COBE}, the BAO signal was measured for the first time in the CMB, detecting small angle anisotropies in the CMB angular power spectrum, confirming the cosmological predictions (\citealt{CMBCl1}; \citealt{CMBCl2}). Later on in 2005, the BAO signal was measured in the Sloan Digital Sky Survey \citep[SDSS; ][]{SDSS} by \cite{LRG} using the spatial correlation function of a spectroscopic subsample of luminous red galaxies (LRG), finding the BAO peak at $r\approx100~h^{-1}$Mpc; and in the 2dF Galaxy Redshift Survey \citep[2dFGRS; ][ 2003]{2dF1} by \cite{Pk2dF} using the power spectrum. Since BAO measurements have proven to be a robust probe for extracting cosmological information, substantial work has been devoted to model and detect the BAO signal in two-point statistics and use it for cosmological constraints (e.g. \citealt{LRGPk}; \citealt{WPer07}; \citealt{Sperg07}; \citealt{BAOBOSS}; \citealt{Ariel09}; \citealt{Reid10}; \citealt{Blake11}; \citealt{Samu13}; \citealt{wedges},b).

There are two important points related to the usual study of LSS using 3D analysis that need to be considered. First, to work in configuration-space, it is necessary to assume a fiducial cosmological model in order to transform the measured angular positions on the sky and redshifts of galaxies into comoving coordinates or distances, a process which could bias the parameter constraints if not treated carefully (see e.g. \citealt{LRG} and \citealt{Ariel09}). Secondly, in order to obtain a precise measurement of either the correlation function or the power spectrum, usually the whole galaxy sample is used to obtain one measurement, typically averaging over a wide redshift range assuming that the measurement at the mean redshift is representative of the entire sample, washing out information on the redshift evolution of the structures.

Even when these two issues are well understood and under control within certain conditions, a simple way to avoid them is by using two-point statistics based only on direct observables, i.e. only angular positions and/or redshifts, such as the angular correlation function $\omega(\theta)$ or the angular power spectrum $C_\ell$. This is done by dividing the sample into redshift bins, or shells, in order to recover information along the line of sight, which otherwise would be lost due to projection effects. In the last few years there have been several papers modelling and analysing large galaxy catalogues using angular two-point statistics. Although most of these focus mainly on photometric-redshift galaxy surveys (\citealt{Crocce11a}, \citealt{Crocce11b}, \citealt{Padm07}, \citealt{Ross11}, \citealt{ESanch}, \citealt{deSimoni13}), this approach has also been applied to spectroscopic-redshift samples (\citealt{Jacobo}, \citealt{Jacobo2}, \citealt{DiDio13}). Here we focus on the cosmological implications of applying this tomographic approach to a BOSS-like spectroscopic-redshift galaxy survey, computing $\omega(\theta)$ in redshift-shells and using this information to obtain constraints on cosmological parameters.

There are three main advantages of this tomographic approach: (i) compared to that of photometric redshifts (photo-z), the higher accuracy of spectroscopic redshifts significantly reduces the overlap between redshift-shells, allowing us to assume that there is no correlation between them due to these uncertainties, and to use thinner shells. Compared to the traditional 3D analysis, (ii) as we already mentioned, by using direct observables we do not need to assume a cosmological model in order to compute spatial separations between galaxies, their angular separations will remain unaffected independent of the cosmological model being tested; (iii) By measuring the angular scale of the BAO peak imprinted on $\omega(\theta)$ at many different redshifts, we are basically measuring the angular diameter distance $D_\rmn{A}(z)$ at several redshifts instead of just one more precise measurement of the average distance $D_V(\bar z)$ at the mean redshift of the sample, giving us more information about the rate at which $D_\rmn{A}$ evolves, putting stronger constraints on the expansion history of the Universe.

This paper is organised as follows. In Section \ref{sec:sample} we describe the mock catalogues that we used, how we configured them in redshift-shells, and what information we expect to extract from the technique discussed in this paper. In Section \ref{sec:Mod} we describe the model we use for the angular correlation function measured in redshift shells and its covariance matrix, and describe the test we perform to assess the ability of these models to extract unbiased cosmological constraints. In Section \ref{sec:BOSS} we describe a synthetic dataset constructed using these models, and present a forecast of the accuracy on cosmological constraints expected from applying this tomographic approach to the final BOSS galaxy catalogue in combination with Planck CMB measurements, comparing this with the constraints that would result from the  combination of the isotropic BAO measurements post-reconstruction on the same catalogue and Planck. We finish with our main conclusions in Section \ref{sec:Conc}.


\section{Angular Correlation Functions in redshift-shells}\label{sec:sample}
In Section \ref{sec:LasDamas} we describe the set of mock catalogues we used for testing our model and the tomographic technique discussed in this paper, and how we configured the catalogues in redshift-shells. In Section \ref{sec:ShellsDa} we explain how to measure cosmological distances using $\omega(\theta)$ in redshift-shells, and what information we expect to extract from these measurements.

\subsection{LasDamas Mock Catalogues}\label{sec:LasDamas}

\begin{table}
\begin{center}
\begin{tabular}{lcc}
\hline\hline
\mbox{Cosmological constant density parameter} & $\Omega_\Lambda$ & $0.75$\\
\mbox{Matter density parameter} &  $\Omega_\rmn{m}$ & $0.25$\\
\mbox{Baryonic density parameter} & $\Omega_\rmn{b}$ & $0.04$\\
\mbox{Dark energy equation of state} & $w_{DE}$ & $-1.0$\\
\mbox{Hubble constant $\left(\mbox{km s}^{-1}\right.$Mpc$\left.^{-1}\right)$} & $H$ & $70$\\
\mbox{Amplitude of density fluctuations} & $\sigma_8$ & $0.8$\\
\mbox{Scalar spectral index} & $n_\rmn{s}$ & $1.0$\\
\\
\hline\\
\mbox{Number of particles} & $N_p$ & $1280^3$\\
\mbox{Box size ($h^{-1}$Mpc)} & $L$ & $2400$\\
\mbox{Particle mass ($10^{10}M_\odot$)} & $M_p$ & $45.73$\\
\mbox{Softening length ($h^{-1}$kpc)} & $\epsilon$ & $53$\\
\hline\hline
\end{tabular}
\caption{Cosmological parameters and specifications of the LasDamas simulation.}\label{tab:cosparam}
\end{center}
\end{table}

\begin{figure*}
 \hskip -0.4cm \includegraphics[scale=0.5]{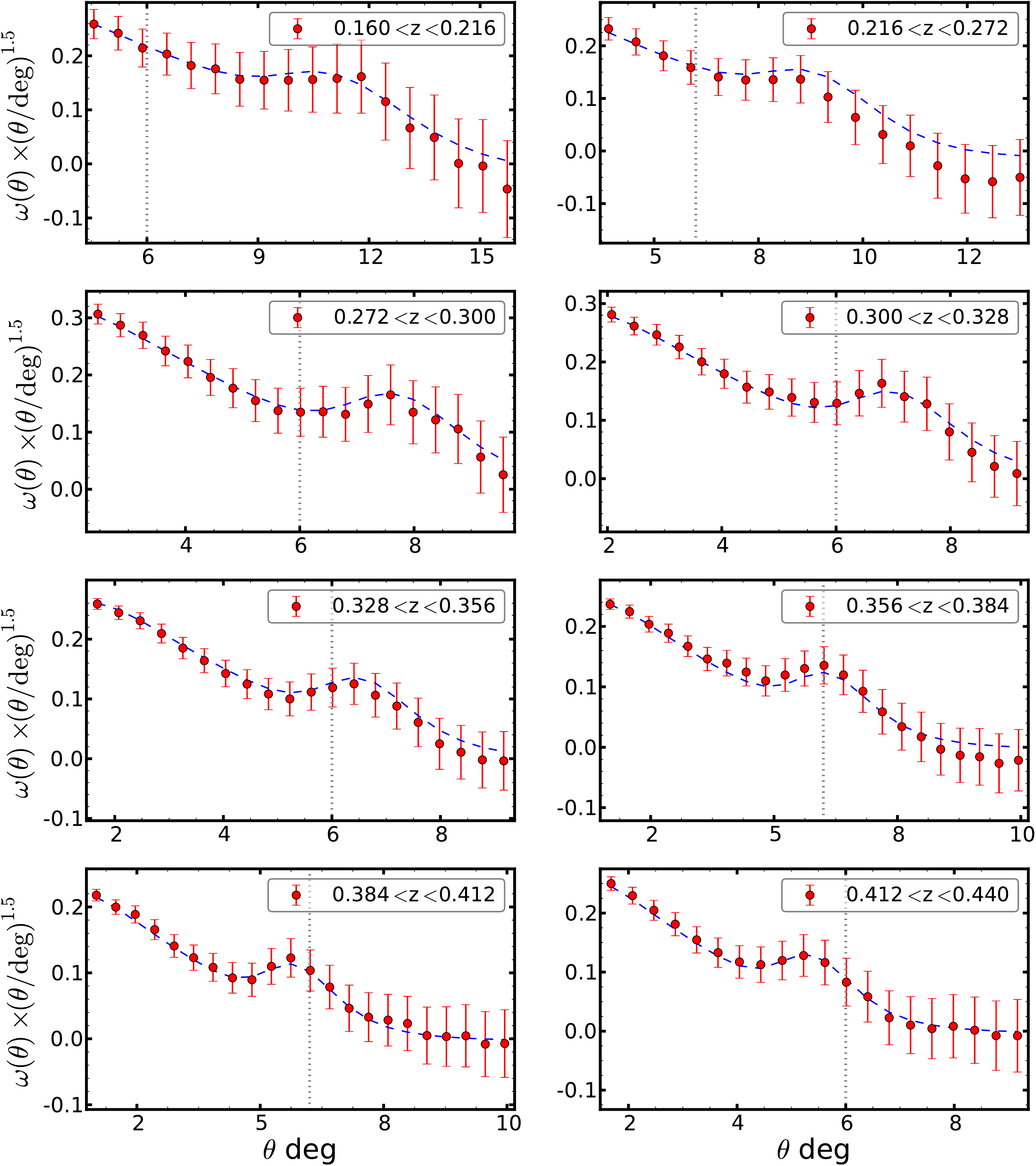}
 \caption{Mean $\omega(\theta)$ measured on the mock catalogues for 8 redshift-shells, amplified by $(\theta/deg)^{1.5}$ to highlight the BAO peak. The errorbars correspond to the error in the mean. The blue dashed lines show the best-fitting model, described in Sec. \ref{sec:modw} and \ref{sec:mod3d}, for the cosmology of LasDamas, which simultaneously reproduces $\omega(\theta)$ for every shell. The vertical dotted line is a reference located at $6$ deg., drawn to show how the BAO peak moves relative to a fixed scale depending on the redshift.}
 \label{fig:shells}
\end{figure*}

We used a set of $160$ spectroscopic luminous red galaxies (LRGs) mock catalogues from LasDamas\footnote{http://lss.phy.vanderbilt.edu/lasdamas} \citep{McB}, constructed from a set of 40 dark-matter only N-body simulations, all of them following the same $\Lambda$CDM cosmological model and using the same initial power spectrum but a different random seed. The specifications of these simulations are outlined in Table~\ref{tab:cosparam}. From each realization, a halo catalogue is extracted using a friends-of-friends algorithm \citep[FoF;][]{Dav85}, and populated with mock galaxies following a halo occupation distribution (HOD; \citealt{P&S2000}, \citealt{B&W02})in order to reproduce the SDSS DR7 \citep{DR7} clustering signal. Each realisation provides $4$ catalogues without overlap, reproducing the SDSS DR7 geometry (northern Galactic cap only), containing an average of $91137$ galaxies per catalogue within the redshift range $[0.16,0.44]$, and including redshift-space distortions (RSD) from peculiar velocities. These catalogues, and the corresponding random field (which contains $50$ times more objects than one of these catalogues) needed to measure the correlation functions, were modified to follow the radial number density $n(z)$ of the SDSS DR7 LRGs (see Fig. 1 in \citealt{FM12}).

\begin{figure}
 \includegraphics[scale=0.55]{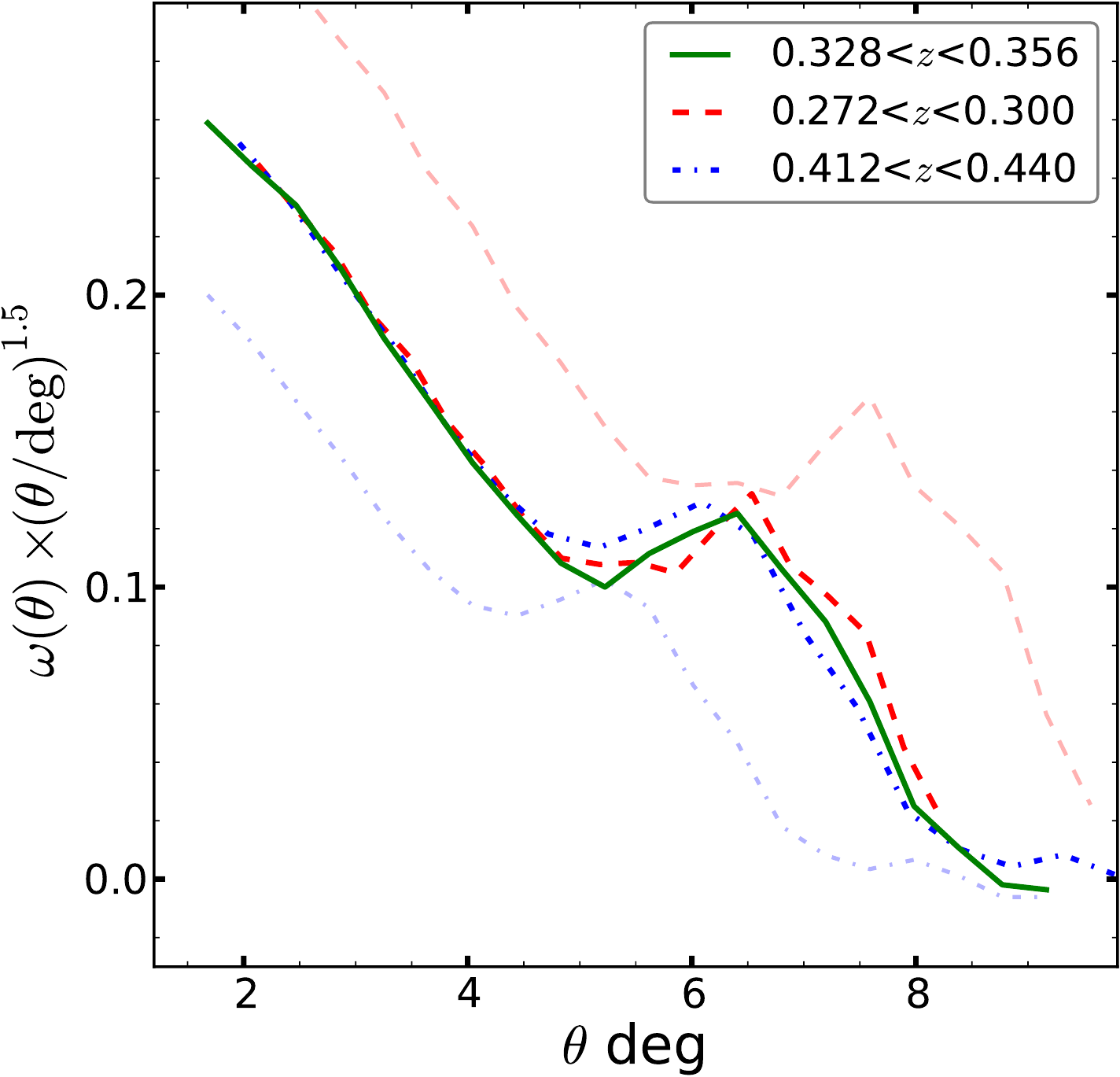}
 \caption{The mean $\omega(\theta)$ measured on LasDamas for three different shells, amplified by $(\theta/deg)^{1.5}$. Two of them have been rescaled following eq. (\ref{eq:alphawth}) (dashed and dash-dotted lines) using the third one as reference (solid line), from their original position (faint-colour version). }
 \label{fig:DaCorr}
\end{figure}

We divided each mock catalogue and the random field into redshift-shells to perform our analysis. Thicker shells lower the signal of the BAO peak, because it is projected over an increasingly wide range of angular scales given by the deeper sample. Thinner shells increase the BAO signal, but decrease the projected number density and therefore decrease the accuracy of the measurements, while increasing the correlation between shells due to RSD effects and the clustering itself. Using a spectroscopic-redshift sample, any overlap between redshift-shells due to redshift uncertainties can be safely neglected as long as their width is much larger than these uncertainties. We tested a number of configurations in order to estimate the optimal redshift bin size, considering the strength of the BAO signal and the uncertainty in measuring $\omega(\theta)$. For simplicity we ignored any correlation between shells, but, as discussed by \citet{Jacobo} and \citet{DiDio13}, cross-correlations should add extra information. The final configuration for LasDamas consists of $8$ shells: $2$ low redshift shells of $\Delta z = 0.056$ covering the redshift range $[0.16,0.272]$, and $6$ higher redshift shells of $\Delta z = 0.028$ covering the redshift range $[0.272,0.44]$.

Using the estimator of \citet{LS93}, we computed the angular correlation function $\omega(\theta)$ in every shell of each mock catalogue and used these measurements to compute the mean $\omega(\theta)$ of each shell and to estimate its associated covariance matrix. These measurements only depend on direct observables (angular positions and redshifts) and do not require the assumption of a fiducial cosmological model to be computed and thus will remain invariant when considering the constraints on cosmological parameters. Fig. \ref{fig:shells} shows the mean $\omega(\theta)$ measured from the 8 shells, amplified by $\theta^{1.5}$ in order to highlight the BAO feature, and where the errorbars correspond to the error in the mean. The dashed lines show the best-fitting model (described in Sections \ref{sec:modw} and \ref{sec:mod3d}) for the cosmology of LasDamas, which simultaneously reproduces $\omega(\theta)$ for every shell.

\subsection{Measuring distances using $\bomega(\btheta)$ in redshift-shells}\label{sec:ShellsDa}

If we look again carefully at Fig. \ref{fig:shells}, it can be seen that the BAO peak in $\omega(\theta)$ is located at different angular scales depending on the redshift shell, i.e. depending on the distance to that shell; this is the key feature that we want to exploit. Let us say that we are only measuring the angular position $\theta_s$ of the BAO peak, then for a given redshift $z_i$ we have
\begin{equation}
 \theta_s(z_i)=r_\rmn{s}(z_d)/D_\rmn{A}(z_i),\label{eq:thsound}
\end{equation}
where $r_s(z_d)$ is the sound horizon at the drag redshift, and $D_\rmn{A}$ is the angular diameter distance given by
\begin{equation}
 D_\rmn{A}(z) = \frac{r(z)}{(1+z)},\label{eq:DA}
\end{equation}
where $r(z)$ is the comoving distance to redshift $z$. Using the fact that the sound horizon corresponds to a fixed scale, in linear theory we can relate its angular scale as $\theta_s(z_i)=\alpha_{ij}\theta_s(z_j)$, where $\alpha_{ij}$ is defined as
\begin{equation}
 \alpha_{ij}\equiv\frac{D_\rmn{A}(z_j)}{D_\rmn{A}(z_i)}.\label{alphaij}
\end{equation}
Then, we can extend this relation to the angular correlation function of two different shells as
\begin{equation}
\omega(\theta,z_i)\simeq\omega(\alpha_{ij}\theta, z_j).\label{eq:alphawth} 
\end{equation}
In Fig. \ref{fig:DaCorr} we show the mean $\omega(\theta)$ measured in three different redshift-shells of LasDamas, where two of them have been rescaled using as reference the third one following eq. (\ref{eq:alphawth}), computing $D_\rmn{A}$ at their mean redshift. The error bars have been omitted for clarity. It can be seen that they match remarkably well on applying the simple relation in eq. (\ref{eq:alphawth}), despite the fact that there are some small differences in their shape due to non-linear evolution of the density field and RSD, which are discussed in Section \ref{sec:Mod}.

The technique discussed in this paper is based on the following idea: if we have $N$ measurements of $\omega(\theta)$ in different redshift shells, in practice we have $N-1$ measurements of $D_\rmn{A}(z_i)/D_\rmn{A}(z_j)$ for $i\neq j$, constraining the rate at which the angular diameter distance can evolve over the redshift range being tested.

\section[]{Modelling $\bomega(\btheta)$ and its covariance matrix} \label{sec:Mod}

Here we describe our model of the two-point angular correlation function used to extract information from the full shape of $\omega(\theta)$ without introducing systematic errors, starting in Section \ref{sec:modw} from the description of its analytical model in thin redshift shells and the distortion effects that have to be taken into account, then going on to describe in Section \ref{sec:mod3d} how to include such effects by modelling the anisotropic two-point spatial correlation function. In Section \ref{sec:theocov} we briefly describe the model for the covariance matrix of $\omega(\theta)$ and compare it with the ones measured from the mock catalogues.
\subsection[]{Angular clustering in redshift shells} \label{sec:modw}
The projection of the spatial density fluctuation field along the line of sight, in a certain direction $\mathbf{\hat n}$ in the sky, is given by
\begin{equation}
 \delta(\mathbf{\hat n}) = \int dz \phi(z) \delta(r\mathbf{\hat n}), \label{eq:dnr}
\end{equation}
where $\phi(z)$ is the radial selection function normalised to unity within a redshift-shell, which for this work is defined as
\begin{equation}
 \phi(z) = \frac{\frac{dN_g}{dz}\vartheta(z)}{{\displaystyle\int}dz \ \frac{dN_g}{dz}\vartheta(z)}, \label{eq:phiofz}
\end{equation}
where $\frac{dN_g}{dz}$ is the number of galaxies per unit redshift, and $\vartheta(z)$, in terms of the redshift range of each shell $[z_i,z_f]$, is given by
\begin{equation}
 \vartheta(z)=\left\lbrace \begin{array}{cc}
                            1 & z_i < z < z_f \\
                            0 & \rmn{otherwise}
                           \end{array}
              \right. .
\end{equation}

Similarly, the angular two-point correlation function can be obtained from the projection of its spatial counterpart $\xi$ \citep{Pee73}. That is, 
\begin{equation}
 \omega(\theta) = \int\int dz_1 dz_2 \phi(z_1) \phi(z_2) \xi \left(s \right), \label{eq:wxi1}
\end{equation}
where $s=\sqrt{ r^2(z_1) + r^2(z_2) - 2r(z_1)r(z_2)\cos\theta }$ is the comoving pair separation, and $\theta$ is the angular separation on the sky.

\begin{figure}
 \includegraphics[scale=0.55]{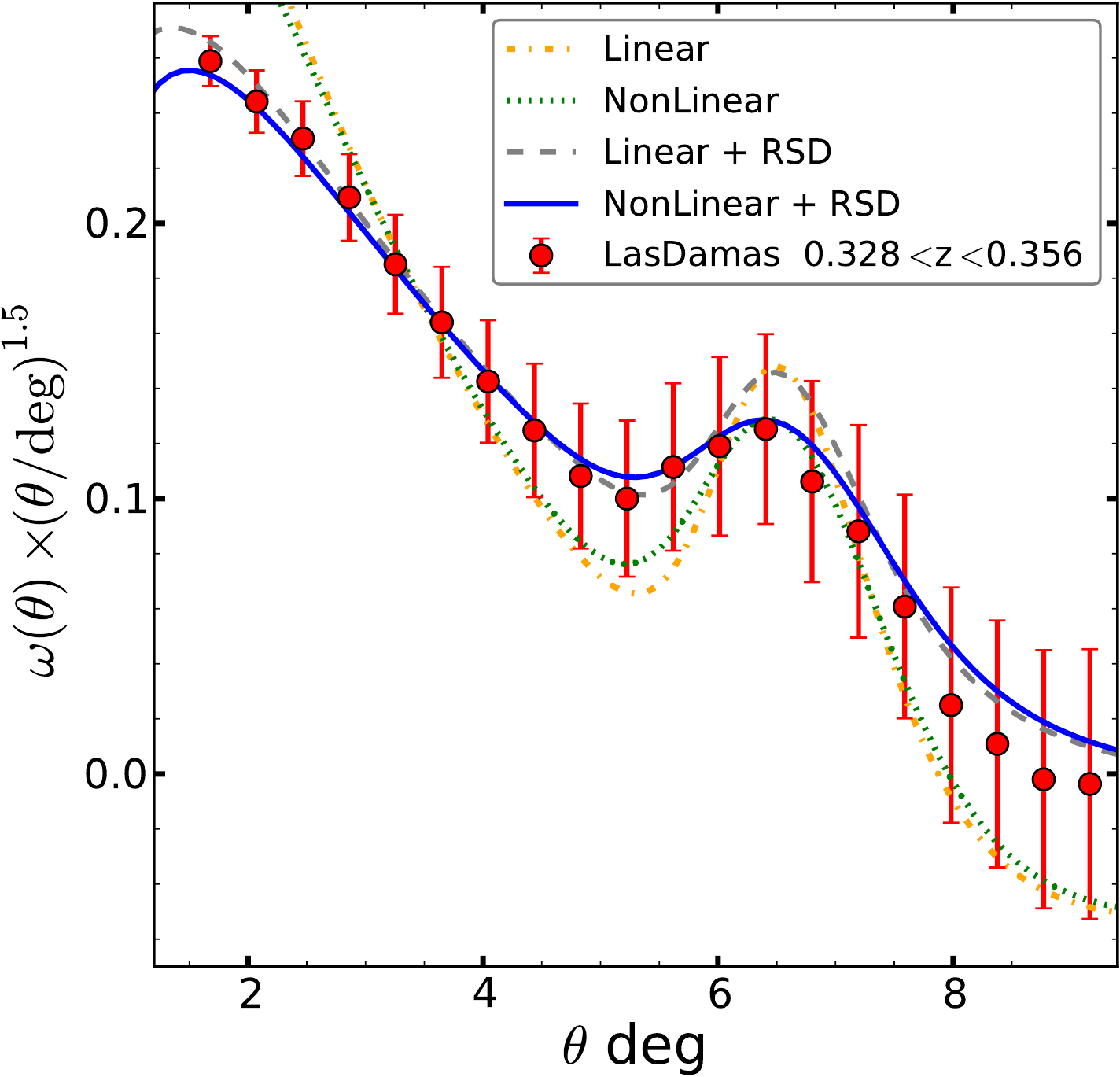}
 \caption{The mean $\omega(\theta)$ amplified by $(\theta/deg)^{1.5}$ for LasDamas (red points) in the redshift-shell $0.328<z<0.356$, and the resulting models obtained including or not non-linear growth and redshift-space distortions (RSD) for the same shell. The green dotted line shows the impact of including non-linear growth effects to the basic linear model (yellow dash-dotted line), while the grey dashed line shows the effect of including RSD in the same linear model. The blue solid line is the final model which includes both effects. Those models that do not include RSD are arbitrarily normalized to match the amplitude of the measurements. The errorbars correspond to the error in the mean.}
 \label{fig:models}
\end{figure}

When working on redshift shells, it is essential to include non-linear effects in the modelling of $\omega(\theta)$ (\citealt{RSDw}; \citealt{Ross11}; \citealt{Fosalba13}). This is shown in Fig. \ref{fig:models}, where different approaches, applying corrections for these effects or not, are compared to the measurements made on the mock catalogues. It can be seen that the RSD corrections have the strongest effects on the full shape of $\omega(\theta)$, but are not enough to describe the damping effects on the BAO peak without including non-linear corrections, which also slightly move the centroid of the peak towards smaller scales. In order to fully describe the shape of $\omega(\theta)$ including these effects, we replaced the spatial correlation function in equation (\ref{eq:wxi1}) by the anisotropic two-dimensional spatial correlation function described in Section \ref{sec:mod3d}. Using this, the model for $\omega(\theta)$ is given by 
\begin{equation}
 \omega(\theta) = \int\int dz_1 dz_2 \phi(z_1) \phi(z_2) \xi \left( s,\mu_s \right), \label{eq:wxi2}
\end{equation}
where $\mu_s$ is the cosine of the angle between the separation vector $\mathbf{s}$ and the line of sight, which in terms of redshift and the angular separation, takes the form of $\mu_s=\frac{r(z_2)-r(z_1)}{s}\cos\left(\frac{\theta}{2}\right)$.

When comparing the model for $\omega(\theta)$ with measurements, it is important to take into account the effect of the binning in $\theta$. Measurements are not done over a single angle $\theta$, but correspond to the average over a bin centred on $\theta$ with a bin-width $\Delta\theta$. In order to avoid systematic effects such as a shift in the BAO peak determination, we consider in our analysis the bin-averaged angular correlation function, evaluated at the bin $\theta_i$, given by

\begin{equation}
 \omega(\theta_i) = \frac{1}{\Delta\Omega_i} \int_{\Delta\Omega_i}d\Omega \ \omega(\theta), \label{eq:binwt}
\end{equation}
where $\Delta\Omega_i$ is the solid angle given by
\begin{equation}
 \Delta\Omega_i = 2\pi\int_{\theta_i-\Delta\theta/2}^{\theta_i+\Delta\theta/2}d\theta^\prime \sin\theta^\prime. \label{eq:DOmega}
\end{equation}

\subsection[]{Anisotropic clustering in redshift-space} \label{sec:mod3d}

Here we describe our model for the anisotropic 3D clustering in redshift-space. In order to take into account the non-linear evolution of the density field, we have based our approach on renormalised perturbation theory \citep[RPT;][]{RPT} to parametrise the non-linear real-space galaxy power spectrum as
\begin{equation}
 P_{NL}(k,z)=b^2\left[P_L(k,z)\rmn{e}^{-(k\sigma_\rmn{v})^2}+A_\rmn{MC}P_{\rmn{MC}}(k,z)\right], \label{eq:Pnlin}
\end{equation}
where the galaxy bias $b$, $\sigma_\rmn{v}$ and $A_\rmn{MC}$ are treated as free parameters, $P_L(k,z)$ is the linear theory power spectrum and $P_{\rmn{MC}}(k,z)$ is given by
\begin{equation}
\begin{split}
 P_{\rmn{MC}}(k,z) = \frac{1}{4\pi^3}\int d^3q & \left[\left| F_2 \left(\mathbf{k}-\mathbf{q},\mathbf{q} \right) \right|^2\right. \\ 
                                               & \left. P_L\left(\left|\mathbf{k}-\mathbf{q}\right|,z\right) P_L(q,z)\right], \label{eq:Pmc}
 \end{split}
\end{equation}
where $F_2(\mathbf{k},\mathbf{q})$ is the standard second order kernel of perturbation theory \citep{Crocce12} given by
\begin{equation}
 F_2(\mathbf{q}_1,\mathbf{q}_2) = \frac{5}{7}+\frac{1}{2}\frac{\mathbf{q}_1\cdot\mathbf{q}_2}{q_1q_2}\left(\frac{q_1}{q_2}+\frac{q_2}{q_1}\right) + \frac{2}{7}\left(\frac{\mathbf{q}_1\cdot\mathbf{q}_2}{q_1q_2}\right)^2. \label{eq:Fq}
\end{equation}

Let us make a break here and consider our specific problem. Unlike the traditional 3D analysis, where it is assumed that evolving quantities such as the galaxy bias $b$ are constant within the sample, in our analysis we need to allow for their evolution. Nevertheless, since each shell is covering a small redshift range, we can neglect the evolution of the density field within a shell, allowing us to evaluate terms such as $b$ and the growth factor $D(z)$ at the mean redshift of the shell $\bar z_\rmn{shell}$. We emphasise that this does not mean that the evolution of the whole sample is negligible, it needs to be considered from shell to shell. With this in mind, starting from the galaxy bias $b$ in the mock catalogues, since theoretical models favour smooth variations in $b$ as a function of redshift for galaxy samples with a fixed selection (\citealt{bCarlton}; \citealt{bKauff}), we assume a linear redshift evolution in which the value of $b$ for a given shell is
\begin{equation}
 b = b_\ast+b^\prime\left(\bar z_\rmn{shell}-z_\rmn{ref}\right),\label{eq:bLD}
\end{equation}
where now $b_\ast$ and $b^\prime$ are our free parameters for the galaxy bias, and $z_\rmn{ref}$ is some reference redshift. We also adopt a redshift evolution for $\sigma_\rmn{v}$ given by
\begin{equation}
 \sigma_\rmn{v} = \sigma^\ast_\rmn{v}\frac{D(\bar z_\rmn{shell})}{D(z_\rmn{ref})},\label{eq:sv}
\end{equation}
where $\sigma^\ast_\rmn{v}$ is now the free parameter. The amplitude of the power spectrum in a given shell is related to that of the reference redshift as
\begin{equation}
 P_L(k,\bar z_\rmn{shell}) = \left(\frac{D(\bar z_\rmn{shell})}{D(z_\rmn{ref})}\right)^2 P_L(k,z_\rmn{ref}).
\end{equation}
We do not assume any redshift evolution for $A_\rmn{MC}$. With these considerations, the set of free parameters of our model, i.e. $\left\lbrace b_\ast,b^\prime,\sigma^\ast_\rmn{v},A_\rmn{MC}\right\rbrace$, are fitted to $z_\rmn{ref}$, and the specific value of $b$ and $\sigma_\rmn{v}$ in each shell is given by eq. (\ref{eq:bLD}) and (\ref{eq:sv}), relating every shell to a single set of values for these free parameters, which in practice means that we are able to simultaneously describe $P_{NL}(k,\bar z_\rmn{shell})$, therefore $\omega(\theta,\bar z_\rmn{shell})$, for every shell.

Back to the anisotropic clustering description, in redshift-space, the two-dimensional power spectrum $P(k,\mu_k)$ can be described by
\begin{equation}
 P(k,\mu_k) = \left( \frac{1}{1+(kf\sigma_\rmn{v}\mu_k)^2} \right)^2 \left( 1+\beta\mu_k^2 \right)^2P_{NL}(k), \label{eq:Pkmu}
\end{equation}
where $f=\left.\frac{\partial\ln D}{\partial\ln a}\right|_{\bar z_\rmn{shell}}\approx\left(\Omega_\rmn{m}(\bar z_\rmn{shell})\right)^\gamma$ is the growth rate factor at $\bar z_\rmn{shell}$, $\gamma$ is the growth index parameter given by the fitting formula in \citet{Linder}, $\beta=f/b$, and $\mu_k$ is the cosine of the angle between the wavevector $\mathbf{k}$ and the line of sight. This parametrization of the redshift-space power spectrum gives a very good description of the {\it Fingers of God} damping effect, where the Lorentzian pre-factor represents a damping function assuming an exponential galaxy velocity distribution function (\citealt{Park94}; \citealt{Cole95}).

\begin{figure}
  \includegraphics[scale=0.55]{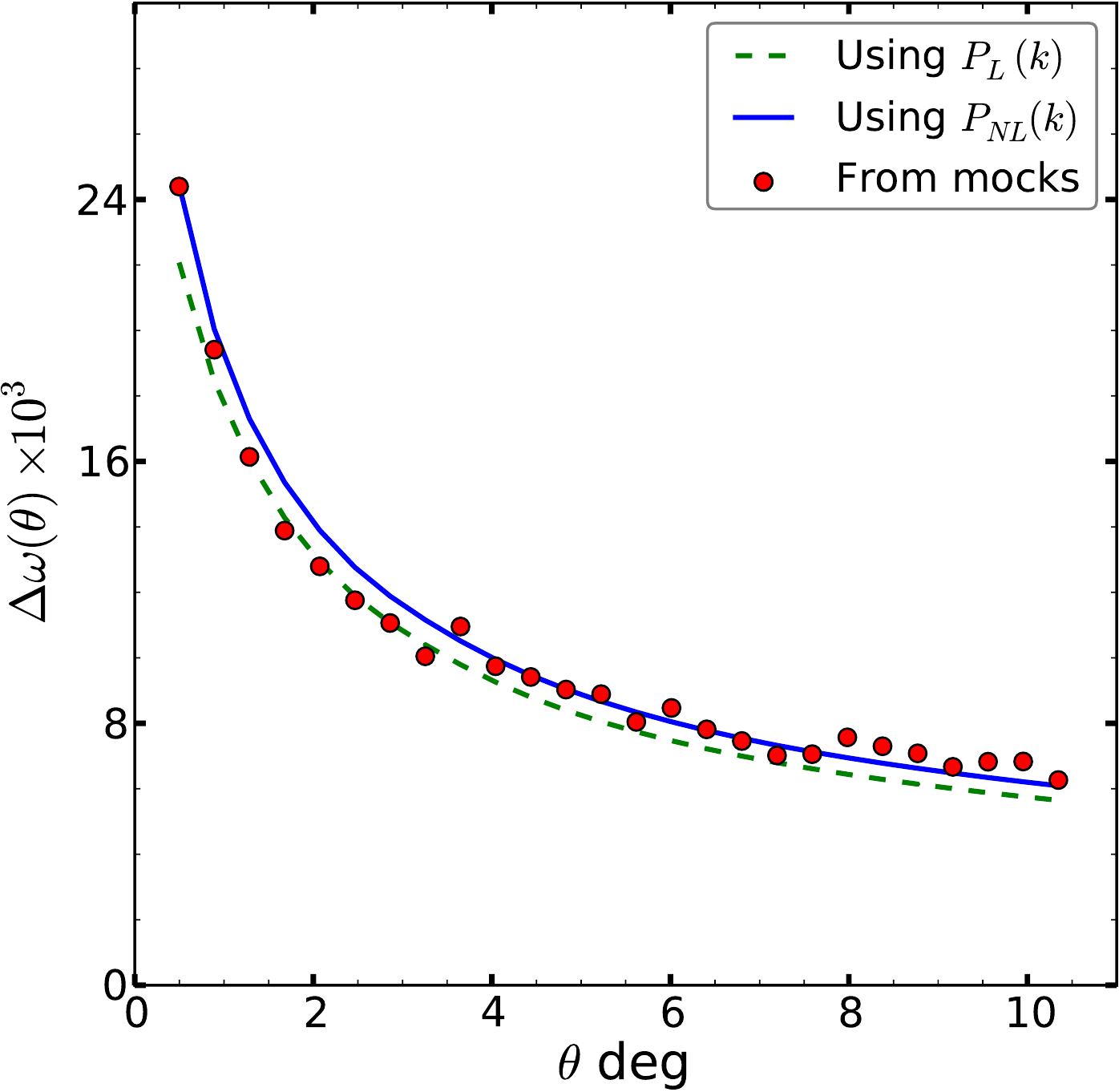}
  \caption{The square root of the variance of $\omega(\theta)$ amplified by $10^3$, as a function of the angular separation, measured on LasDamas in the redshift shell $0.412< z < 0.44$ (red points); and the analytical prediction obtained using both, $P_L(k)$ (green dashed line) and $P_{NL}(k)$ (blue solid line), into the modelling.}
  \label{fig:Cov}
\end{figure}

Following the procedure described by \citet{wedges}, it is convenient to expand the two-dimensional spatial correlation function $\xi \left( s,\mu \right)$ as
\begin{equation}
 \xi \left( s,\mu_s \right) = \sum_{\ell \ \rmn{even}} \xi_{\ell}(s)L_{\ell}(\mu_s), \label{eq:xiexp}
\end{equation}
where $L_{\ell}(\mu_s)$ are the Legendre Polynomials of even $\ell$-th order. Even though in theory this is an expansion over infinite even multipoles, just a few of them have a non-negligible contribution on the scales of interest in this work (see \citealt{wedges}), meaning that, in practice, most of the information is enclosed in the monopole and the quadrupole, and the multipoles of order $\ell\geq4$ can be safely neglected. Then, the expression used to model the spatial anisotropic clustering is given by 
\begin{equation}
 \xi(s,\mu_s) = \xi_0(s) + L_2(\mu_s)\xi_2(s), \label{eq:xi2}
\end{equation}
where $\xi_0(s)$ and $\xi_2(s)$ are the monopole and the quadrupole of $\xi(s)$ respectively. To model these multipoles, we can expand the two-dimensional power spectrum $P(k,\mu_k)$ in a similar way using Legendre polynomials, where each multipole $P_\ell(k)$ can be computed as 
\begin{equation}
 P_\ell(k) = \frac{2\ell+1}{2} \int d\mu_k \ P(k,\mu_k)L_\ell(\mu_k), \label{eq:Pl}
\end{equation}
from which the $\xi_\ell(s)$ multipoles are given by 
\begin{equation}
 \xi_\ell(s)=\frac{i^\ell}{2\pi^2}\int dk \ k^2 P_\ell(k) j_\ell(ks). \label{eq:xil}
\end{equation}
where $j_\ell(x)$ is the spherical Bessel function of $\ell$-th order \citep{Hamilton97}.

These models, for both the power spectrum and the correlation function, have been shown to give a remarkably good description of non-linear evolution and RSD in measurements of both N-body simulations (\citealt{Ariel08}; \citealt{FM10}) and real galaxy samples (\citealt{Ariel09}; \citealt{FM12}; \citealt{Ariel12}; \citealt{wedges},b).

\subsection[]{The Covariance Matrix of $\bomega(\btheta)$}\label{sec:theocov}

\begin{figure*}
 \hskip -0.7cm $\begin{array}{ccc}
   \includegraphics[scale=0.32]{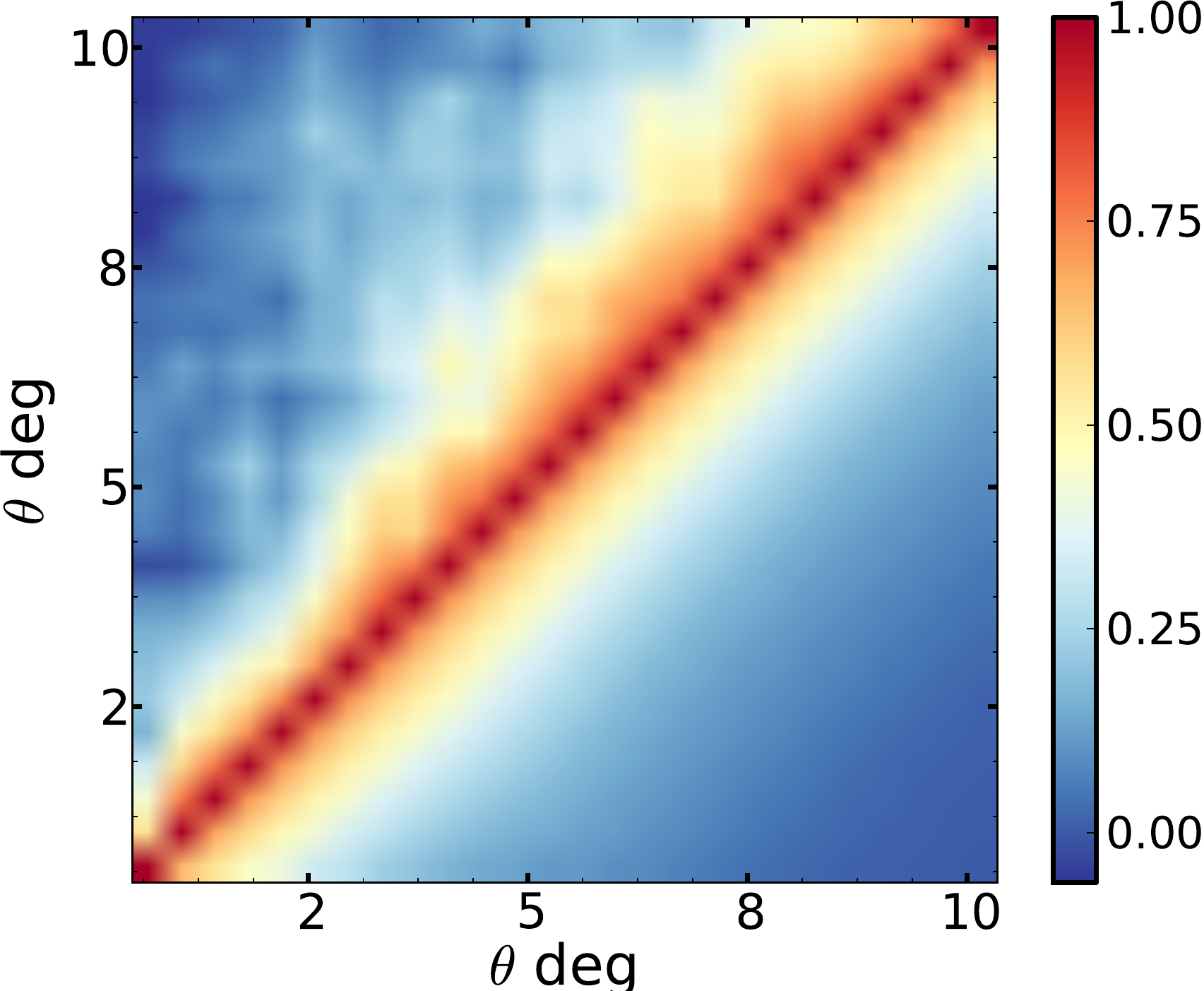} &
   \includegraphics[scale=0.325]{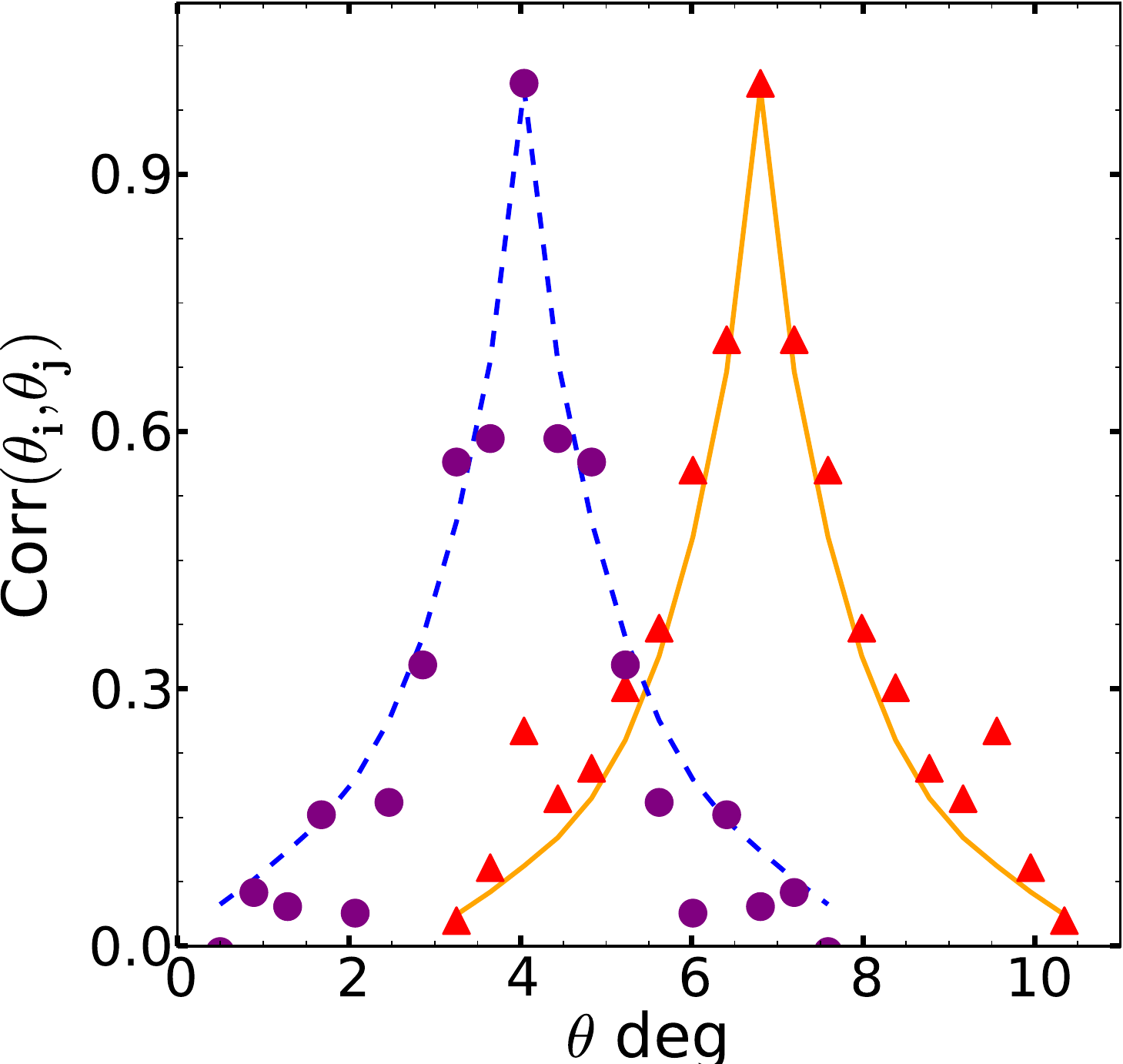} &
   \includegraphics[scale=0.325]{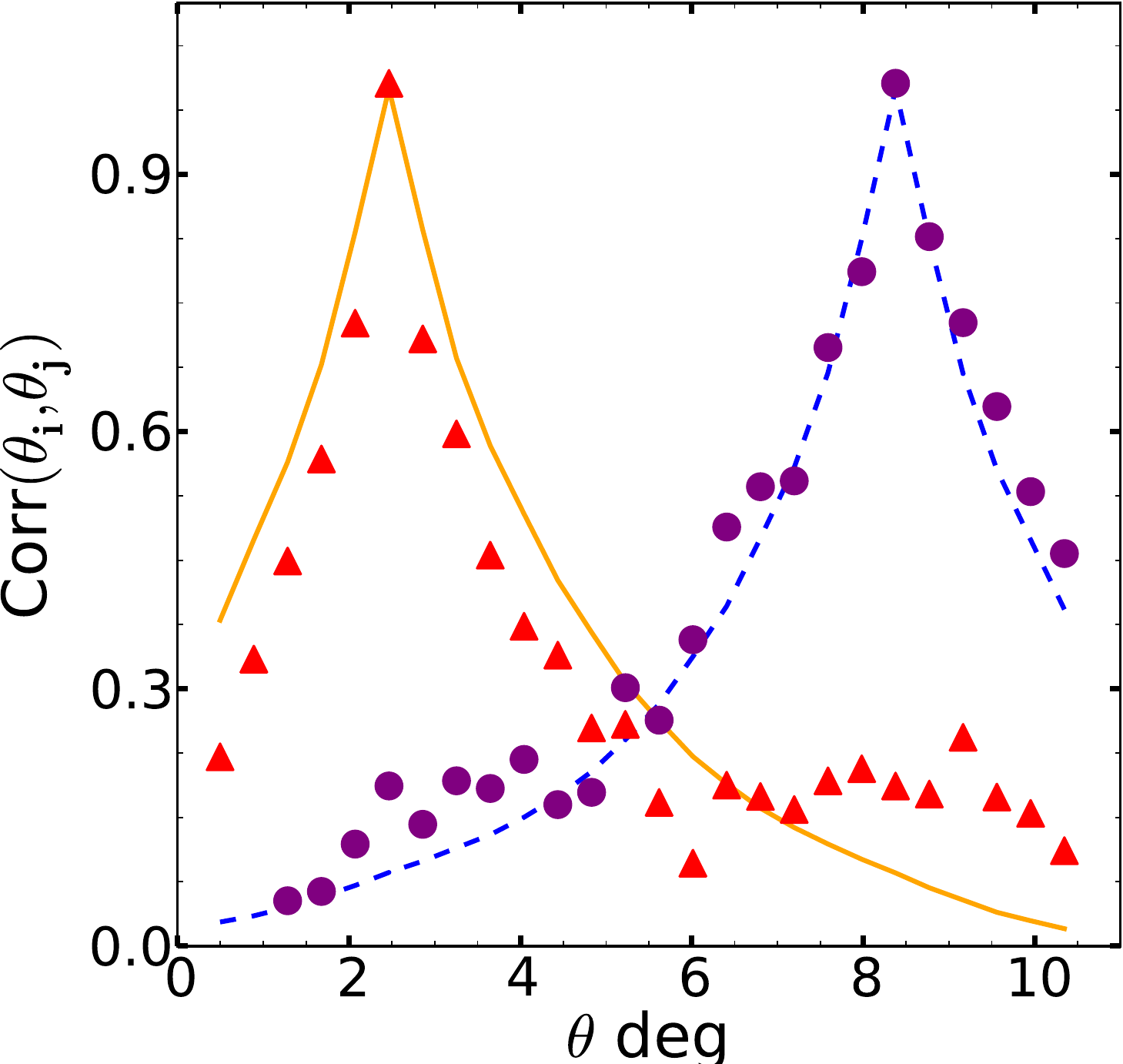}
 \end{array}$
  \caption{Left: correlation matrix of $\omega(\theta)$ measured on LasDamas in the redshift shell $0.412< z < 0.44$ (upper-triangular) and its analytical prediction using $P_{NL}(k)$ (lower-triangular). Centre: two anti-diagonals of the same matrix, where the purple circles and red triangles are the measurements on LasDamas and the dashed blue and solid yellow lines correspond to the analytical matrix, respectively. Right: two horizontal cuts of the same matrix, following the same symbology as the central panel.}
  \label{fig:Corr}
\end{figure*}

Since the set of mock catalogues from LasDamas consists of only 160 realizations, a direct estimation of the full covariance matrix of $\omega(\theta)$ in redshift-shells would be noisy \citep{WPer13}. That is why we use an analytical model instead, following the recipe of \cite{Crocce11a}. Here we briefly describe the more important steps, and refer the reader to their article for a more detailed description. 

The angular galaxy power spectrum $C_\ell$ in redshift-space for a redshift shell is given by
\begin{equation}
 C_\ell = \frac{2}{\pi}b^2D^2(\bar z_\rmn{shell})\int dk \ k^2 P(k) \left(\Psi_\ell(k)+\beta\Psi^r_\ell(k)\right)^2, \label{eq:Cl}
\end{equation}
where $\Psi_\ell$ and $\Psi^r_\ell$ are the real- and redshift-space contributions to the kernel function given by
\begin{equation}
 \Psi_\ell(k) = \int dz \phi(z) j_\ell(kr), \label{eq:psik}
\end{equation}
and 
\begin{equation}
 \begin{split}
  \Psi^r_\ell(k) = & \int dz \phi(z) \left[ \frac{2\ell^2+2\ell-1}{(2\ell+3)(2\ell-1)}j_\ell(kr) \right. \\
                   & -\frac{\ell^2-\ell}{(2\ell-1)(2\ell+1)}j_{\ell-2}(kr) \\
                   & \left.-\frac{(\ell+1)(\ell+2)}{(2\ell+1)(2\ell+3)}j_{\ell+2}(kr) \right]. \label{eq:psirk}
 \end{split}
\end{equation}

Then, the covariance matrix including the shot-noise contribution can be computed as
\begin{equation}
 \rmn{Cov}_{\theta_i\theta_j} = \frac{2}{f_\rmn{sky}}\sum_{\ell\geq2}\frac{2\ell+1}{(4\pi)^2}L_\ell\left(\mu_i\right)L_\ell\left(\mu_j\right)\left(C_\ell+\frac{1}{\bar n}\right)^2, \label{eq:TCov}
\end{equation}
where $\mu_i = \cos\theta_i$, $f_\rmn{sky}$ is the fraction of the sky observed, and $\bar n$ is the number of objects per steradian. 

As well as with $\omega(\theta)$, the covariance matrix is affected by the fact that measurements are done over a bin in $\theta$, reducing the covariance between bins (\citealt{Cohn06}, \citealt{Ariel08}, \citealt{Smith08}). We consider the bin-averaged covariance matrix obtained from averaging over $\Delta\Omega_i$ and $\Delta\Omega_j$. Each of these integrals only affect the Legendre polynomials evaluated at $\cos\theta_i$ and $\cos\theta_j$ respectively. Defining
\begin{equation}
 \begin{split}
  \hat L_\ell(\mu_i) & =  \frac{1}{\Delta\Omega_i} \int_{\Delta\Omega_i} d\Omega L_\ell(\mu_i) \\
                             & =  \frac{2\pi}{\Delta\Omega_i}\frac{1}{2\ell+1}\left[ L_{\ell+1}\left(\mu_i^+\right) + L_{\ell-1}\left(\mu_i^-\right) \right. \\
                             & \hskip 0.4cm    \left. - L_{\ell+1}\left(\mu_i^-\right) - L_{\ell-1}\left(\mu_i^+\right) \right], \label{eq:binLell}
 \end{split}
\end{equation}
where $\mu_i^\pm = \cos(\theta_i\pm\Delta\theta/2)$, the bin-averaged covariance matrix (which as before, we will keep denoting just as $\rmn{Cov}_{\theta_i\theta_j}$) is then given by
\begin{equation}
 \rmn{Cov}_{\theta_i\theta_j} = \frac{2}{f_\rmn{sky}}\sum_{\ell\geq2}\frac{2\ell+1}{(4\pi)^2}\hat L_\ell\left(\mu_i\right)\hat L_\ell\left(\mu_j\right)\left(C_\ell+\frac{1}{\bar n}\right)^2. \label{eq:binTCov}
\end{equation}

We tested this model for the covariance matrix, using in equation (\ref{eq:Cl}) both $P_L(k)$ and $P_{NL}(k)$ with the best-fitting values of $\lbrace b_\ast$,$b^\prime\rbrace$, and $\lbrace\sigma_\rmn{v}^\ast$,$A_\rmn{MC}\rbrace$ when needed, for the cosmology of LasDamas, and compared the results with the estimated matrix from the mock catalogues. Fig. \ref{fig:Cov} shows the square root of the diagonal elements of the covariance matrix for the shell within $0.412< z < 0.44$, which is the dispersion of $\omega(\theta)$ in this shell, estimated from the mock catalogues (red points), the prediction using the linear power spectrum (green dashed line), and the prediction using the non-linear power spectrum (blue solid line). It can be seen that both approaches, specially the non-linear one, give a very good description of the variance of the angular correlation function for the scales in which we are interested. Hereafter we will only use the non-linear approach.

The left panel of Fig. \ref{fig:Corr} shows the reduced covariance matrix, or correlation matrix, defined as 
\begin{equation}
 \rmn{Corr}_{\theta_i\theta_j}=\frac{\rmn{Cov}_{\theta_i\theta_j}}{\sqrt{\rmn{Cov}_{\theta_i\theta_i}\rmn{Cov}_{\theta_j\theta_j}}},
\end{equation}
for the same shell as Fig. \ref{fig:Cov}, where the upper-triangular part is the estimation from the mock catalogues and the lower-triangular part corresponds to the theoretical model. The central panel shows two anti-diagonals of the correlation matrix estimated from the mock catalogues (points), and of the predicted matrix (solid lines). The same symbols apply for the right panel, where two horizontal cuts of these matrices are shown.

We computed the theoretical matrix for every shell using $P_{NL}(k)$, and used them to test our technique.

\subsection{Testing the model for $\bomega(\btheta)$}\label{sec:test}

In order to test the model for the angular correlation function, we implemented a Markov chain Monte Carlo (MCMC) analysis taking the same $\Lambda$CDM cosmology of LasDamas (see Table \ref{tab:cosparam}) and exploring the following parameter space:
\begin{equation}
 \mathbb{P}_\rmn{test} \equiv \lbrace w_{DE}, b_\ast, b^\prime, \sigma_\rmn{v}^\ast, A_\rmn{MC} \rbrace, \label{eq:Ptest}
\end{equation}
where $w_{DE}$ is the constant dark energy equation of state parameter, and the rest are the free parameters of our model for $\omega(\theta)$. We estimate the likelihood function as $\mathcal{L}(\mathbf{P}_\rmn{test})\propto\exp\left(-\chi^2(\mathbf{P}_\rmn{test})/2\right)$, where
\begin{equation}
 \chi^2\left(\mathbf{P}_\rmn{test}\right) = \sum_\rmn{shells} (\mathbf{M}_i-\mathbf{D}_i)^T\mathbf{\widehat{Cov}}_i^{-1}(\mathbf{M}_i-\mathbf{D}_i),\label{eq:shchi2}
\end{equation}
$\mathbf{P}_\rmn{test}$ is a vector with the parameter values, $\mathbf{M}_i$ is the model of the shell $i$ given $\mathbf{P}_\rmn{test}$, $\mathbf{D}_i$ is the mean $\omega(\theta)$ measured in the shell $i$, and $\mathbf{\widehat{Cov}}_i$ is the corresponding covariance matrix for the same shell divided by $\sqrt{N_\rmn{mocks}}$, which represents the covariance matrix for a volume equal to the total volume of the ensemble, allowing us to detect any bias in the constraints. To compute the models for $\omega(\theta)$, the linear power spectrum $P_L(k)$ is calculated using {\sc camb} \citep{Camb}.

The goal here is to test if we are able to recover the correct value of $w_{DE}$ using this model and the measurements made on LasDamas. Since our model does not have any free parameter to adjust the position of the BAO peak on $\omega(\theta)$, and moreover, it reproduces this angular scale simultaneously for every shell, recovering the correct value of $w_{DE}$ basically means that we are able to correctly measure the distance to every single redshift-shell, describing the expansion history of the Universe.

\begin{figure}
 \includegraphics[scale=0.55]{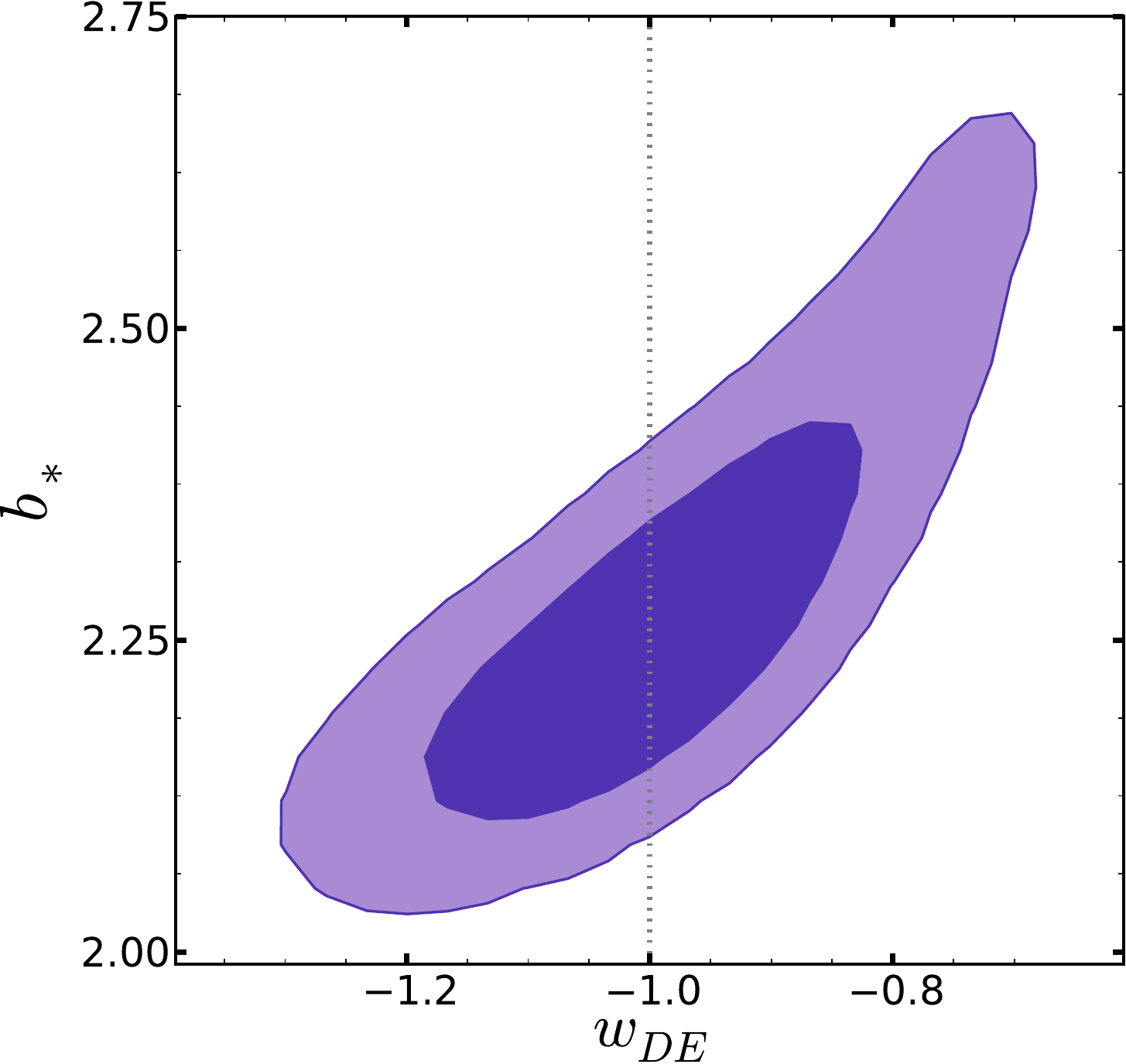}
 \caption{The marginalised 68 and 95 per cent confidence levels in the $w_{DE}-b_\ast$ plane for our test. Here we find $w_{DE}=-0.99\pm0.12$ in excellent agreement with the correct value used to construct the mock catalogues (dotted line).}
  \label{fig:w-b}
\end{figure}

Fig. \ref{fig:w-b} shows the resulting marginalised constraints in the $w_{DE}-b_\ast$ plane, where the contours are the 68 and 95 per cent confidence levels. For this test we found $w_{DE}=-0.99\pm0.12$, which is in excellent agreement with the true value of LasDamas, showing that this technique is able to extract unbiased constraints on $w_{DE}$.

\section{BOSS Forecast}\label{sec:BOSS}

We tested the implications of applying this technique to the final SDSS-III BOSS catalogue (DR12), in combination with Planck, for three different flat cosmological models, and compared this with what would result from the combination of Planck and isotropic BAO measurements post-reconstruction on BOSS (CMASS and LOWZ). To do so, we characterised the BOSS catalogue by assuming the best fit of the base $\Lambda$CDM model from Planck plus WMAP polarization (WP) as our true cosmology \citep{planck}, an area in the sky of $10000$ deg$^2$, a constant $n(z)=3\times10^{-4}h^3$Mpc$^{-3}$, and a galaxy bias based on \citet{bBOSS} given by

\begin{equation}
 b = 1+\frac{\left(b_0-1\right)}{D(\bar z_\rmn{shell})}\label{eq:bBOSS},
\end{equation}
which describes its redshift evolution for the CMASS sample. Also, since the effect of massive neutrinos is not negligible in the Hubble expansion rate $H(a)$, we adopted the exact treatment in \citet{WMAP7} given by
\begin{equation}
 \begin{split}
  H(a) = & H_0\left( \frac{\Omega_\rmn{b}+\Omega_\rmn{cdm}}{a^3} +\frac{\Omega_\gamma}{a^4}\left( 1 + 0.2271N_\rmn{eff}f(m_\nu a/T_{\nu0})\right) \right. \\ 
         &\left. + \frac{\Omega_\rmn{k}}{a^2} + \frac{\Omega_\Lambda}{a^{3(1+w_{DE}(a))}}\right), \label{eq:HmassNu}
 \end{split}
\end{equation}
where $a$ is the scale factor, $m_\nu a/T_{\nu0} = \left(1.87\times10^5/(1+z)\right) \Omega_\nu h^2$, the photon density parameter is $\Omega_\gamma = 2.469\times10^{-5}h^{-2}$ for $T_\rmn{cmb}=2.725\rmn{K}$, and $f(y)$ can be approximated by the fitting formula
\begin{equation}
 f(y) \approx \left( 1+\left( Ay \right)^p \right)^\frac{1}{p},
\end{equation}
where $A=180\zeta(3)/(7\pi^4)$, $\zeta(3)\simeq1.202$ is the Riemann zeta function, and $p=1.83$.

\begin{figure*}
 \vskip 0.5cm \includegraphics[scale=0.55]{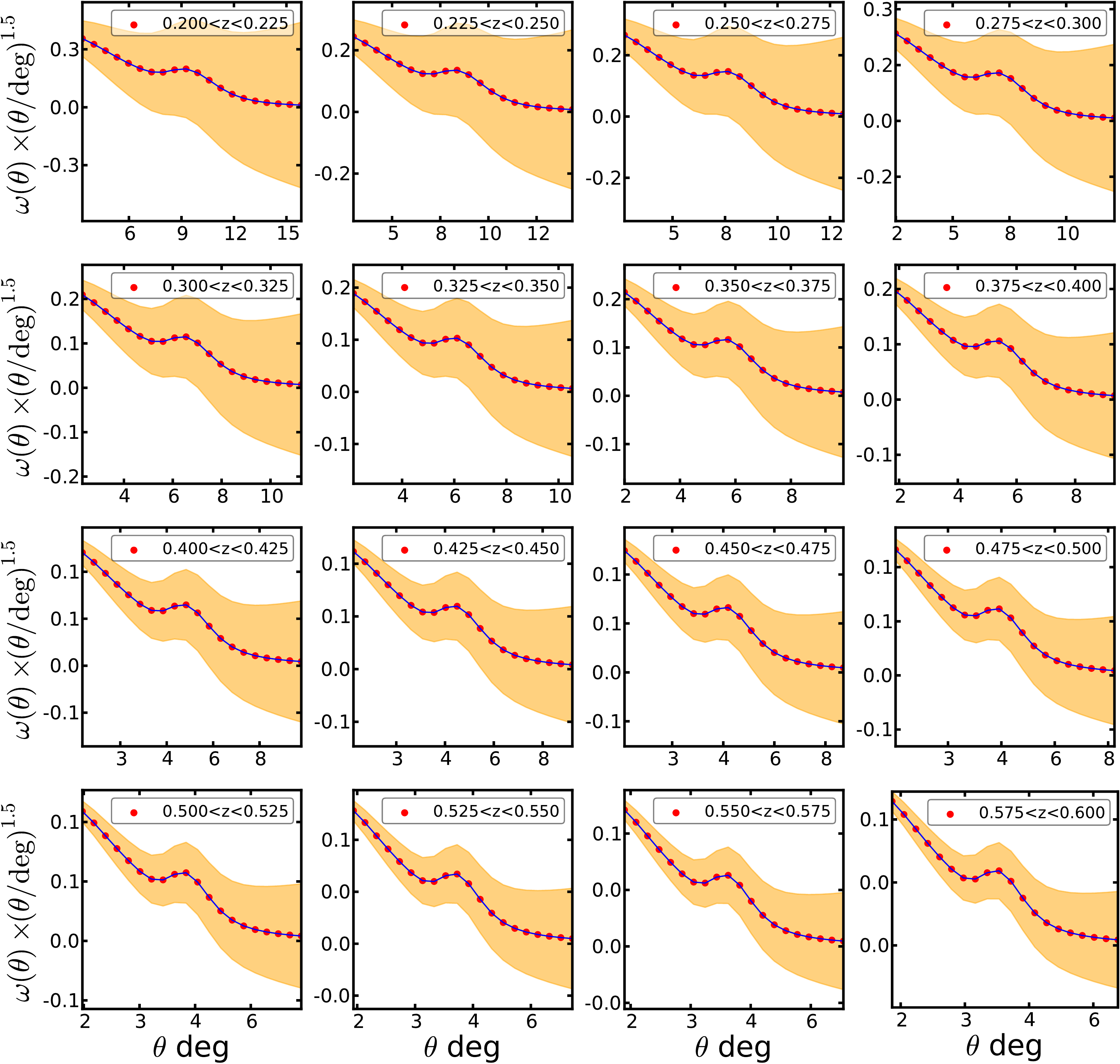}
 \caption{Synthetic dataset constructed with the models of $\omega(\theta)$ and its covariance matrix, taking the best-fit $\Lambda$CDM cosmology from Planck. It consists of 16 redshift shells of $\Delta z=0.025$ within the redshift range $0.2<z<0.6$. We used this dataset to forecast the results of combining Planck and the technique discussed in this paper applied to the final BOSS. We have characterised the BOSS catalogue by assuming an area in the sky of $10000$ deg$^2$, a constant $n(z)=3\times10^{-4}h^3$Mpc$^{-3}$, and a galaxy bias based on \citet{bBOSS}.}
  \label{fig:16shells}
\end{figure*}

Using the model for $\omega(\theta)$ and its covariance matrix described in Sections \ref{sec:modw} and \ref{sec:theocov} respectively, we constructed a synthetic dataset consisting of $16$ shells of width $\Delta z=0.025$, ranging from $z=0.2$ up to $z=0.6$. The fiducial values for the free parameters of the model are $b_0=1.55$, $\sigma_\rmn{v}^\ast=4.29$, and $A_\rmn{MC}=1.5$. The result of this synthetic dataset can be seen in Fig. \ref{fig:16shells}.

For the CMB data we used the distance priors based on \citet{PDistP} which summarises the CMB information from Planck in a set of parameters and its covariance matrix, where we have included the spectral amplitude $A_\rmn{s}$. The resulting set is 
\begin{equation}
\mathbb{P}_\rmn{CMB} \equiv \lbrace \ell_A,R,\omega_\rmn{b},A_\rmn{s},n_\rmn{s}\rbrace, \label{eq:Pcmb}
\end{equation}
where in practice the first 2 parameters, the CMB angular scale $\ell_A$ and the shift parameter $R$, are derived from the other explored parameters in our analysis, which are described below in this Section, following the equations in \citet{PDistP}.

To reproduce the isotropic BAO measurements post-reconstruction on BOSS, for our fiducial cosmology we took the ratio
\begin{equation}
 x(z_m) = \frac{D_\rmn{V}(z_m)}{r_\rmn{s}(z_d)},\label{eq:BAO}
\end{equation}
at $z_m^1=0.32$ with an error of $2\%$ for LOWZ and at $z_m^2=0.57$ with an error of $1\%$ for CMASS \citep{BAOBOSS}, where $D_\rmn{V}(z)$ is the average distance from the mean redshift $z_m$ given by
\begin{equation}
 D_\rmn{V}(z) = \left( (1+z)^2 D_\rmn{A}^2 \frac{cz}{H(z)} \right)^\frac{1}{3}. \label{eq:Dv}
\end{equation}

With these three ingredients we performed a MCMC analysis with the aim of forecasting the accuracy expected, constraining cosmological parameters, from applying this technique to the final BOSS catalogue. The base model for the analysis is the flat $\Lambda$CDM model, where baryons, cold dark matter (CDM) and dark energy (vacuum energy or a cosmological constant $\Lambda$) are the main contributors to the total energy of the Universe; and with Gaussian, adiabatic primordial scalar density fluctuations following a power-law spectrum for the amplitudes in Fourier space. This model can be characterised by the following parameter space:
\begin{equation}
\mathbb{P}_{\Lambda\rmn{CDM}} \equiv \lbrace \omega_\rmn{b},\omega_\rmn{cdm},\omega_\rmn{de},A_\rmn{s},n_\rmn{s}\rbrace, \label{eq:LCDM}
\end{equation}
where $\omega_\rmn{b}$, $\omega_\rmn{cdm}$ and $\omega_\rmn{de}$ are the baryon, cold dark matter and dark energy densities respectively; here $\omega_X\equiv\Omega_Xh^2$. The primordial power spectrum is characterised by its amplitude $A_\rmn{s}$ and its spectral index $n_\rmn{s}$, both defined at the pivot wavenumber $k_p=0.05$ Mpc$^{-1}$. We also extended the base model allowing variations in the dark energy equation of state parameter $w_{DE}$, considering $w_{DE}(a)=w_0$ constant in time, and also a time dependence given by the standard linear parametrization of \citet{w0wa01} and \citet{w0wa03}
\begin{equation}
 w_{DE}(a) = w_0 + w_a(1-a). \label{eq:w0wa}
\end{equation}
Then, the two cases of the extended parameter space are
\begin{equation}
\mathbb{P}_{w\rmn{CDM}} \equiv \lbrace \omega_\rmn{b},\omega_\rmn{cdm},\omega_\rmn{de},A_\rmn{s},n_\rmn{s},w_0[,w_a]\rbrace, \label{eq:wCDM}
\end{equation}
where $[,w_a]$ denotes the variation (or not) of $w_a$. It is also necessary to include the free parameters of our model to the sets in eq. (\ref{eq:LCDM}) and (\ref{eq:wCDM}), in order to compare the constraints obtained from the use of $\omega(\theta)$ and the other dataset combinations. We consider a case where we use the correct bias evolution in eq. (\ref{eq:bBOSS}), treating $b_0$ as a free parameter, giving us three nuisance parameters for our model ($b_0$, $A_\rmn{MC}$ and $\sigma_\rmn{v}^\ast$), and a second case where we do not assume that we know the functional form of the bias evolution, using the linear model in eq. (\ref{eq:bLD}) giving us four nuisance parameters for our model for $\omega(\theta)$ ($b_\ast$, $b^\prime$, $A_\rmn{MC}$ and $\sigma_\rmn{v}^\ast$). We do not consider more flexible parametrizations for the bias evolution, since it is expected that the galaxy bias has a smooth variation as a function of redshift, specially for passively evolving galaxy populations such as LRGs (\citealt{bCarlton}, \citealt{bKauff}, \citealt{Almeida08}).

We estimate the likelihoods as in Section \ref{sec:test}, computing the $\chi^2$ for $\omega(\theta)$ as in eq. (\ref{eq:shchi2}) using the full covariance matrix, and where the argument vector now is $\mathbf{P}$ which has values of the parameter-space corresponding to the cosmology being tested. The $\chi^2$ for the CMB is given by
\begin{equation}
 \chi^2_\rmn{cmb} (\mathbf{P}) = \left(\mathbf{V}_\rmn{cmb}-\mathbf{V}^f_\rmn{cmb}\right)^T\mathbf{Cov}_\rmn{cmb}^{-1}\left(\mathbf{V}_\rmn{cmb}-\mathbf{V}^f_\rmn{cmb}\right),
 \label{eq:cmbchi2}
\end{equation}
where $\mathbf{V}_\rmn{cmb}$ is a vector with the values of $\mathbb{P}_\rmn{CMB}$ as a function of $\mathbf{P}$, $\mathbf{V}^f_\rmn{cmb}$ is the vector with the correct values for our fiducial cosmology, and $\mathbf{Cov}_\rmn{cmb}$ is the covariance matrix for these CMB parameters. For the BAOs, we calculate the $\chi^2$ as follows,
\begin{equation}
 \chi^2_\rmn{bao} = \left( \frac{x(z_m^1)-x^f(z_m^1)}{\sigma_{z_m^1}} \right)^2 + \left( \frac{x(z_m^2)-x^f(z_m^2)}{\sigma_{z_m^2}}  \right)^2
\end{equation}
where $x(z_m^i)$ is the expression in eq. (\ref{eq:BAO}) at $z_m^i$ as a function of $\mathbf{P}$, $x^f(z_m^i)$ is the same expression at $z_m^i$ evaluated in our fiducial cosmology, and $\sigma_{z_m^i}$ is the assumed error for the BAO measurement at $z_m^i$.

\begin{figure}
 \includegraphics[scale=0.445]{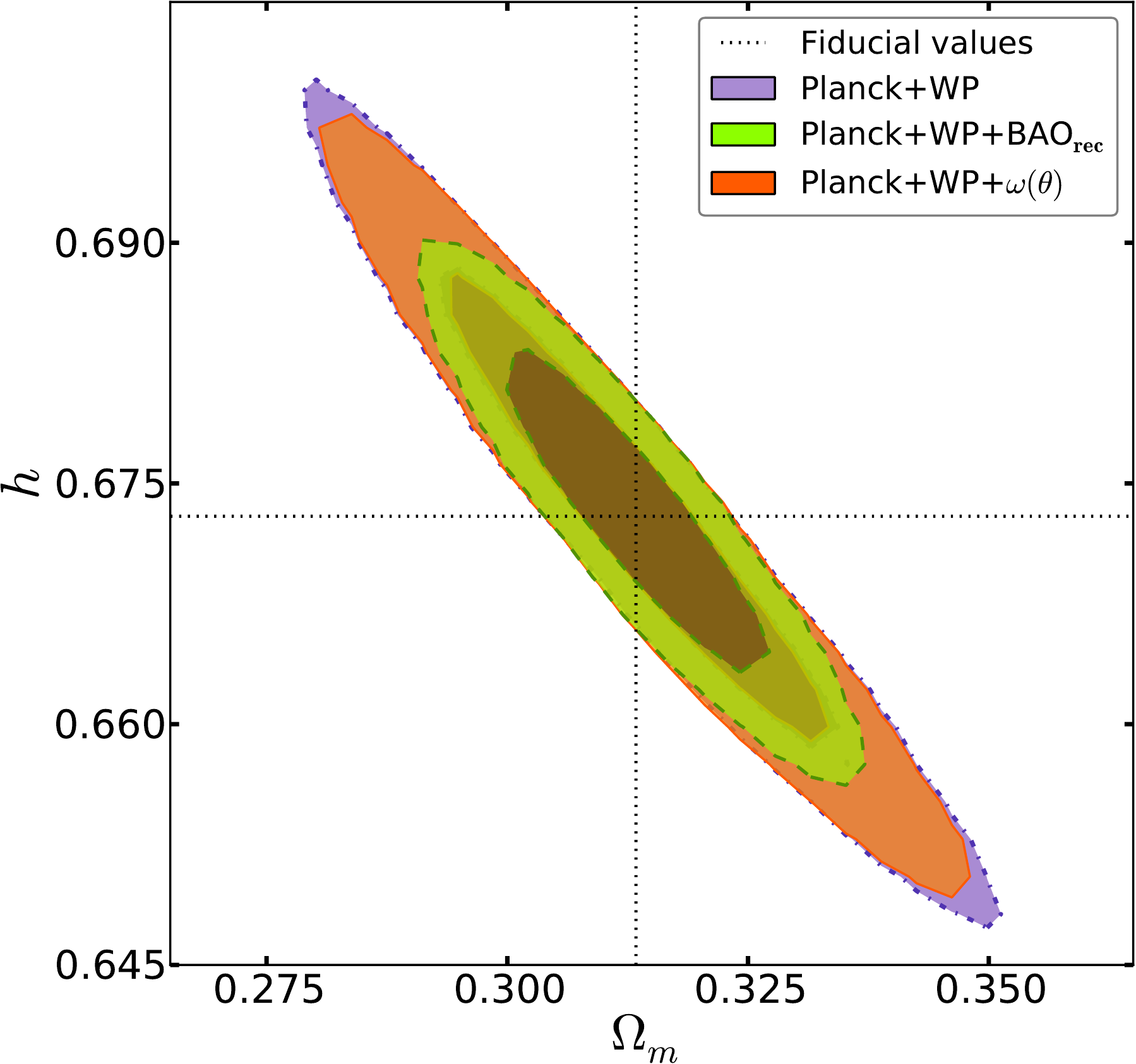}
 \caption{The marginalised 68 and 95 per cent confidence levels in the $\Omega_\rmn{m}-h$ plane for the base $\Lambda$CDM model case. The dash-dotted lines (purple contours) correspond to the constraints derived from the use of Planck+WP only. The dashed lines (green contours) are the constraints obtained by combining Planck+WP and BAO measurements post-reconstruction, while the solid lines (orange contours) are those derived from the combination of Planck+WP and $\omega(\theta)$ without any reconstruction. The dotted lines correspond to the fiducial values assumed to make our forecast.}
  \label{fig:Om-h}
\end{figure}

\begin{figure}
 \includegraphics[scale=0.445]{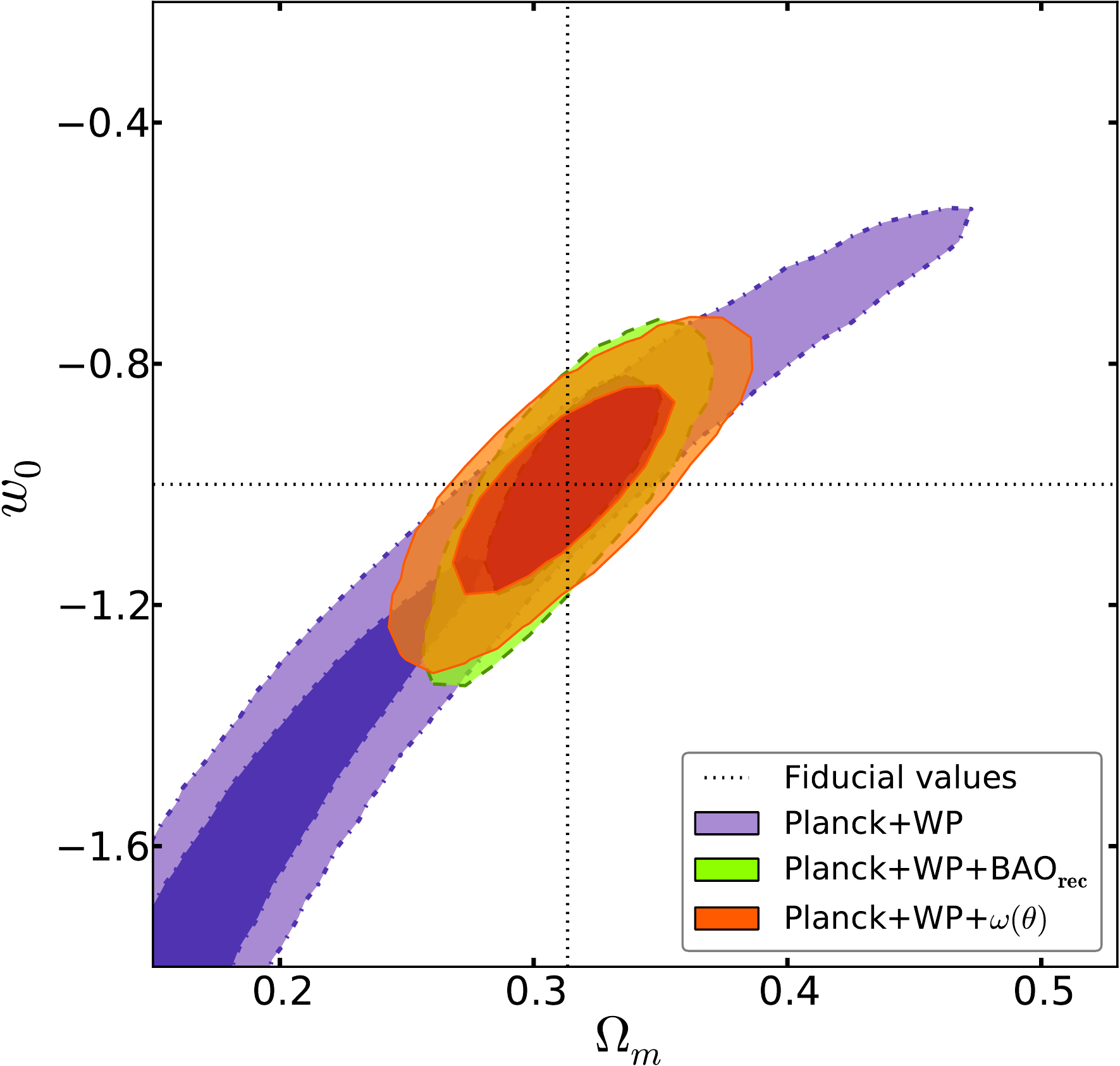}
 \caption{The marginalised 68 and 95 per cent confidence levels on the $\Omega_\rmn{m}-w_0$ plane for the extended $\Lambda$CDM model case with constant $w_{DE}=w_0$. The dash-dotted lines (purple contours) correspond to the constraints derived from the use of Planck+WP only. The dashed lines (green contours) are the constraints obtained by combining Planck+WP and BAO measurements post-reconstruction, while the solid lines (orange contours) are those derived from the combination of Planck+WP and $\omega(\theta)$ without any reconstruction. The dotted lines correspond to the fiducial values assumed to make our forecast.}
  \label{fig:Om-w}
\end{figure}

In the case of the base $\Lambda$CDM model, Fig. \ref{fig:Om-h} shows the marginalised constraints in the $\Omega_\rmn{m}-h$ plane for the different combinations of datasets, where the contours correspond to the 68 and 95 per cent confidence levels. From the combination of Planck+WP and $\omega(\theta)$ on the final BOSS  we find a mean value of $\Omega_\rmn{m}=0.314\pm0.013$ ($68\%$C.L.) and $h=0.673\pm0.010$ ($68\%$C.L.) for the correct bias model, with negligible variation for the linear bias model ($<3\%$), in remarkable agreement with the fiducial cosmology, tightening the constraints derived from the CMB only. Although, it can be seen that, in this case, the combination of Planck+WP and BAO measurements post-reconstruction on BOSS does somewhat better. Nevertheless, once we allow $w_{DE}$ to take a constant value different from $-1$, the constraints from combining Planck+WP and $\omega(\theta)$ are now as good as those obtained from the combination of Planck+WP and BAO measurements post-reconstruction. This can be seen in Fig. \ref{fig:Om-w}, where the contours correspond to the marginalised constraints in the $\Omega_\rmn{m}-w_0$ plane showing the 68 and 95 per cent confidence levels. In this case we find a mean value of $\Omega_\rmn{m}=0.311\pm0.028$ ($68\%$C.L.) and $w_0=-1.00\pm0.11$ ($68\%$C.L.) for the correct bias model, and $\Omega_\rmn{m}=0.308\pm0.032$ ($68\%$C.L.) and $w_0=-1.01\pm0.14$ ($68\%$C.L.) for the linear bias model, again in excellent agreement with our true cosmology.

\begin{figure}
 \includegraphics[scale=0.445]{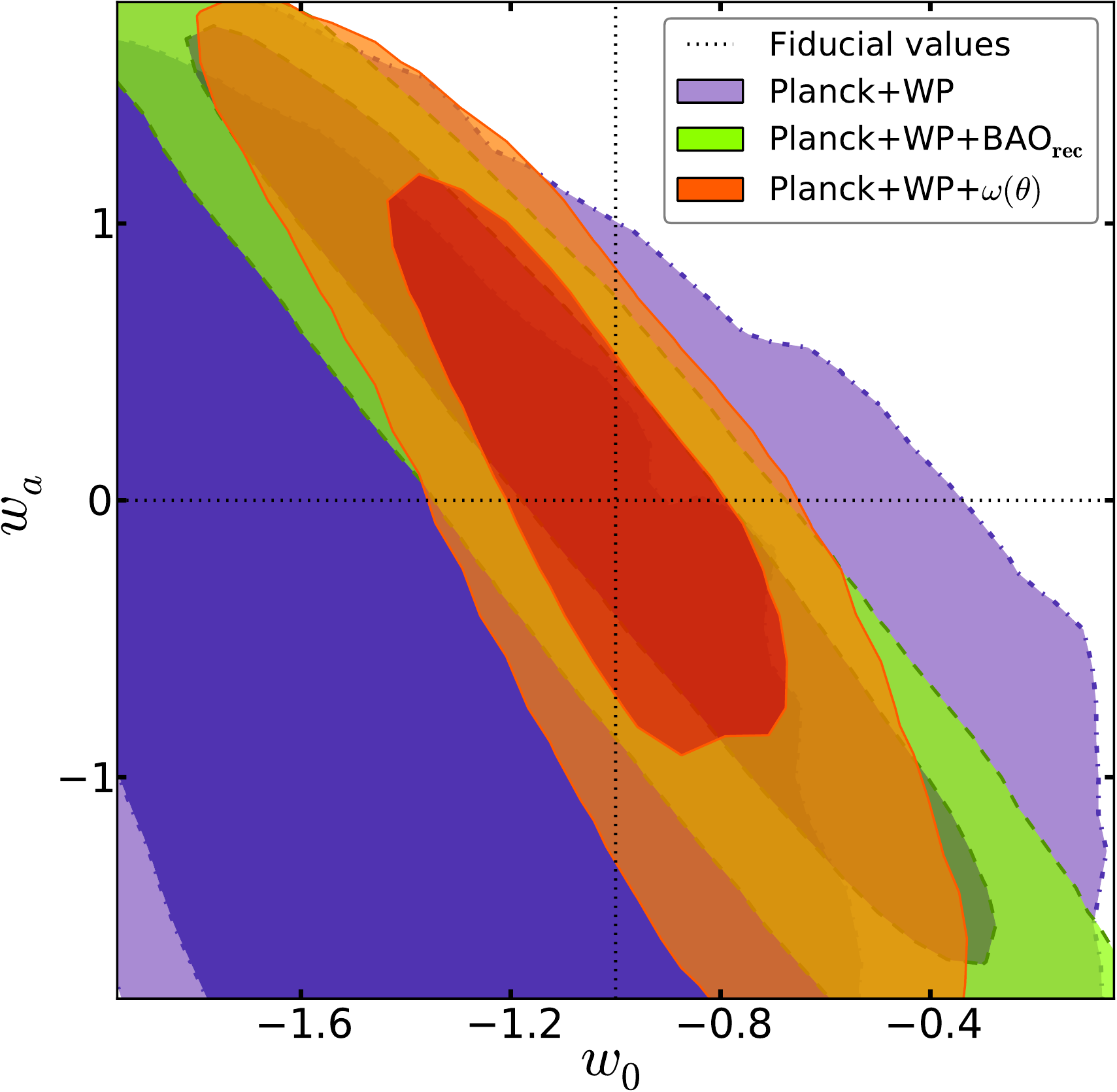}
 \caption{The marginalised 68 and 95 per cent confidence levels in the $w_0-w_a$ plane for the extended $\Lambda$CDM model case, with a time-dependent $w_{DE}$ parametrised as in eq. (\ref{eq:w0wa}). The dash-dotted lines (purple contours) correspond to the constraints derived from the use of Planck+WP only. The dashed lines (green contours) are the constraints obtained by combining Planck+WP and BAO measurements post-reconstruction, while the solid lines (orange contours) are those derived from the combination of Planck+WP and $\omega(\theta)$ without any reconstruction. The dotted lines correspond to the fiducial values assumed to make our forecast.}
  \label{fig:w0-wa}
\end{figure}

If we now allow $w_{DE}$ to vary over time following the parametrization given in eq. (\ref{eq:w0wa}), the constraints obtained from the combination of Planck+WP and $\omega(\theta)$ in this case are much more accurate than those obtained from combining Planck+WP and BAOs. Fig. \ref{fig:w0-wa} shows the 68 and 95 per cent confidence level marginalised constraints in the $w_0-w_a$ plane for the different combinations of datasets, where this accuracy improvement can be seen. From Planck+WP+$\omega(\theta)$ we find a mean value of $w_0=-1.03\pm0.25$ ($68\%$C.L.) and $w_a=0.008_{-0.74}^{+0.76}$ ($68\%$C.L.) for the correct bias model, and $w_0=-1.05\pm.33$ ($68\%$C.L.) and $w_a=0.015_{-0.89}^{+0.91}$ ($68\%$C.L.) for the linear bias model, again in perfect agreement with our fiducial cosmology just like the two previous cases. To quantify the constraints obtained in this case using different dataset combinations, we used the Figure-of-Merit (FoM) defined as (\citealt{FoM1}; \citealt{FoM2})
\begin{equation}
 \rmn{FoM}=\rmn{det}\left[\rmn{Cov}(w_0,w_a)\right]^{-1/2}, \label{eq:FoM}
\end{equation}
where $\rmn{Cov}(w_0,w_a)$ is the $2\times2$ covariance matrix of $w_0$ and $w_a$. The higher the FoM, the more accurate are the constraints made by a particular dataset combination. From the combination of Planck+WP and BAOs the FoM$=9.17$, while from the combination of Planck+WP+$\omega(\theta)$ we obtain a value of $10.54$, increasing the FoM by $15\%$ for the correct bias model. Using the linear bias model, we obtain a FoM of $8.24$, $10\%$ lower compared to the BAO post-reconstruction technique.

What can be concluded from these tests is: (i) The choice of different models for the galaxy bias evolution has an impact on the accuracy that we can constrain cosmological parameters, but a sensible choice can still result in unbiased constraints; (ii)  The more freedom we allow for the expansion history in a given model, the better performance this tomographic approach has compared to the traditional BAO technique. This can be explained mainly by two factors. First, while BAOs only take into account the position of the BAO feature measuring the quantity in eq. (\ref{eq:BAO}), the full shape of the correlation function is also sensitive to other combinations of cosmological parameters, such as $\omega_\rmn{b}$ and $\omega_\rmn{cdm}$. Second, as we mentioned in Section \ref{sec:ShellsDa}, measuring $\omega(\theta)$ in several redshift bins basically gives several measurements of $\theta_s(z)$, constraining the ratio at which the angular diameter distance can evolve over the redshift range being tested. Then if we include the extra information of the full shape of $\omega(\theta)$ mentioned before, we end up with a very powerful tool to probe the expansion history of the Universe.

\section{Conclusions}\label{sec:Conc}

We tested the implications of applying a tomographic approach to a spectroscopic-redshift galaxy survey through measuring the two-point angular correlation function $\omega(\theta)$ in thin redshift shells, avoiding the need to assume a fiducial cosmological model in order to transform measured angular positions and redshifts into comoving distances, as it is the case in the traditional 3D analysis. In principle, this technique, as it is presented in this paper, can be also applied to narrow-band photometric surveys (e.g. PAU survey\footnote{http://www.pausurvey.org/}) without any further consideration, since the accuracy in their photometric-redshifts determination is expected to be $\sim0.0035(1+z)$, but we have not studied this case here.

We first contrasted the predictions of the model for $\omega(\theta)$ and its covariance matrix, described in Section \ref{sec:Mod}, against measurements made on a set of 160 mock catalogues, and tested its ability to recover the correct value of the dark energy equation of state parameter $w_{DE}$ used to construct these catalogues. For simplicity, we did not include cross-correlations between shells in our analysis, although these should add extra information. Our modelling includes effects such as redshift-space distortions and non-linear evolution of the density fluctuations, showing that these effects are completely necessary in order to correctly reproduce the full shape of $\omega(\theta)$. This technique results in an unbiased way to extract cosmological information.

Next, we made a forecast of the accuracy in cosmological constraints expected from applying this technique to the final BOSS galaxy catalogue in combination with the Planck CMB results in three different flat cosmological models, and compared it with what would result from combining Planck and isotropic BAO measurements post-reconstruction on the same galaxy catalogue. To do so, we chose the best-fit of the base $\Lambda$CDM cosmology from Planck as our true cosmology and characterised the BOSS catalogue assuming an area of $10000$ deg$^2$, a constant $n(z)=3\times10^{-4}h^3$Mpc$^{-3}$, the redshift range $0.2<z<0.6$, and the galaxy bias model in eq. (\ref{eq:bBOSS}). Using the model for the angular correlation function and its covariance matrix, we constructed a synthetic dataset consisting of 16 measurements of $\omega(\theta)$ on the same number of redshift-shells covering the whole redshift range. We also computed the CMB likelihood using distance priors for Planck, and reproduce the likelihood obtained from BAO measurements post-reconstruction on BOSS using eq. (\ref{eq:BAO}) and assuming an error of $2\%$ for LOWZ and $1\%$ for CMASS. 

Across this analysis, we used two different models for the galaxy bias evolution, in one case using the correct model used to construct the synthetic dataset, and in the other case using the simpler model in eq. (\ref{eq:bLD}). The different choices showed no biasing constraining cosmological parameters, but different accuracies. The first cosmological model tested was the basic $\Lambda$CDM, which resulted in tighter constraints for the combination of Planck and BAO measurements compared to the combination of Planck and $\omega(\theta)$ measurements. Although, for the other two models tested, where we extended the base model allowing $w_{DE}$ to deviate from its fiducial value of $-1$, being a constant in one case, and allowing a time-dependence in the other, we found that the more freedom we give to $w_{DE}$ the better performance our technique has, resulting in a comparable accuracy when constraining a constant $w_{DE}$ with respect to Planck+BAOs, and up to $15\%$ higher FoM compared to the combination of Planck and BAO measurements in the case of a time-dependent dark energy equation of state, showing that this tomographic approach to analyse the galaxy clustering is able to put strong constraints in the expansion history of the Universe.

\section*{Acknowledgments}
SSA would like to thank Tatiana Tapia and Jan Grieb for useful comments and discutions. SSA and AGS acknowledge support by the Trans-regional Collaborative Research Centre TRR33 {\it The Dark Universe} of the Deutsche Forgschungsgemeinschaft (DFG). SSA and NDP acknowledge support by the European Commission's Framework Programme 7, through the Marie Curie International Research Staff Exchange Scheme LACEGAL  (PIRSES-GA​-2010-2692​64). NDP acknowledges support from BASAL CATA PFB-06 and Fondecyt \#1110328. CMB was supported by the Science and Technology Facilities Council [grant number ST/F001166/1]. We thank the LasDamas project for releasing publicly the mock catalogues. Some of the results in this work have been derived using the {\sc HEALPix} \citep{healpix} package.


\label{lastpage}

\end{document}